# Object Classification by means of Multi-Feature Concept Learning in a Multi Expert-Agent System

Nima Mirbakhsh
CS Department, IASBS
nmirbakhsh@iasbs.ac.ir

Arman Didandeh
CS Department, IASBS
a_didandeh@iasbs.ac.ir

Mohsen Afsharchi
CS Department, IASBS
afsharchim@iasbs.ac.ir

*Abstract*— **Classification of some objects in classes of related concepts is an essential and even breathtaking task in many applications. A solution is discussed here based on Multi-Agent systems. A kernel of some expert agents in several classes is to consult a central agent decide among the classification problem of a certain object. This kernel is moderated with the center agent, trying to manage the querying agents for any decision problem by means of a data-header like feature set. Agents have cooperation among concepts related to the classes of this classification decision-making; and may affect on each others' results on a certain query object in a multi-agent learning approach. This leads to an online feature learning via the consulting trend. The performance is discussed to be much better in comparison to some other prior trends while system's message passing overload id decreased to less agents and the expertism helps the performance and operability of system win the comparison.**

*Keywords-component; Agent, Feature, Concept Learning, Expertism*

## I. INTRODUCTION

In this part, we represent some definitions to clarify the expressions we are going to use in the following parts of this paper:

### AGENTS, MAS AND EXPERTISM

Fortunately or unfortunately, there are very several definitions for an Agent like [8] or [9]. To meet the term in a more general look, an agent is a tool that carries out some task or tasks on behalf of a human. Intelligent agents have additional domain knowledge that enables them to carry out their tasks even when the parameters of the task change or when unexpected situations arise. Many intelligent agents are able to learn, from their own performance, from other agents, from the user, or from the environment in which they are situated. An intelligent Agent has properties as: Autonomy, Ability to Learn, Cooperation &many other ones [4]. Giving a more formal definition, an agent is defined in [5].

A Multi Agent System is a system consisting of several, not-the-same-type agents which are in collaboration with eachother to do a complicated task, usually with some constraints like time and recourses. The MAS is studied for many reasons and by so many pioneers of AI, esp. for Agent-Studying, Decision Making and also Expert Systems. The most important obstacle to be met in perusal of MAS is the trade-off between performance-based and goal-based nature of agents gathering together to model the MAS, where usually a trade-off seems to be undeniable.

"The goal of Multi Agent Systems' research is to find methods that allow us to build complex systems composed of autonomous agents who, while operating on local knowledge and possessing only limited abilities, are nonetheless capable of enacting the desired global behaviors." [6]

Regarding an Expert-Agent, we usually mean an agent which is able to decide about a specific topic so much better than some others agents. This ability may be gained with agent collaboration in a MAS, with learning among some dataset or using any other way of decision optimization. If we have a MAS in which agent are to be experts in their own area of deciding problems, then we will have a Multi Expert-Agent System.

### CONCEPTS, OBJECTS AND FEATURES

As we look around into the world we live, it is available to understand several things and differentiate them from others. This task of differentiation is because of that these being compared things are under some sort of experiment to show their specific characteristics as a react in response to some query-like act. These responses are generated due to the characteristics which are known under the set of "Features". It is clear that the sets we are mentioning here are not disjoint, but have several in common among eachother.

The THINGS which are carrying these feature sets as a collection of properties are called "Objects". Objects are exactly what we see in the world around. But what is making the difference between these objects to generate the responses? Several objects may carry several feature sets which have nothing in common or may share some or even all features in their feature sets. Besides the set of all features that can introduce the objects to us, one more important matter to be reminded is the values that these features can take and are taking in comparison to other objects in common. So an object is defined with the set of all features carrying in its feature set; and also with the specific values that these features are taking among the set of all values they can be instantiated with.

But what is happening when some objects seem the same or at least very likely to be the same when they are being studied? Of course that these objects are near in the set of features and also they have feature values that make them come about eachother. When we have such groups of objects that while being looked, they remind us of some other objects of their like, we are mentioning a more abstract element of knowledge representation called the "Concept". Concepts are the collection of all objects that are usually carrying feature set with most proximity. The more these collections of objects share in their feature sets – and also in the set of values they can take-, the more we call them an integrated unique concept. It is clear for sure that the level of proximity between the members of a Concept-family can be modeled under the standpoint of Object Orientation.

### Concept Learning

A very important hindrance about communication in Multi Agent systems is always put on where some agents are willing to deal with each other about some concepts of which they know nothing or even are not sure to have unanimous ideas about. This makes us deal with the problem of *concept learning*. Several attempts was made on the process of concept learning where some agent learns some concept from one or a set of peer agents into the system they are collaborating in together. A good example of "New Concept Learning" could be met in [1].

We here address the attempts made in [3] while it has some idea of which we believe is having a lot common to our approach. In [3], Afsharchi et al. brings to front the idea of non-unanimous concepts for which they allow the definition of a concept fluctuate between two boundaries of *core* and *periphery* objects. We will later talk more on our approach and the analogy of it with [3] in base ideas.

### II. Approach Overview

Assume that we have some set of main classes with a cardinality of M, let's say $\{C_1, C_2, …, C_M\}$. Any query, consisting of an object to decide about, could go mapped to one or some of these main classes to be classified into our system. This is so similar to many application-based previous tasks, of which we mention here [2] and [10]. In both mentioned works, objects are assumed to be text in a web page. [2] uses a rule-based system for the task of classification and the rules are extracted from known Machine Learning methods with a good support of initial datasets. [10] has this task done with the help of a Naïve-Bayes classifier in a hierarchical structure.

Here we introduce M agents, let to name $\{Ag_1, Ag_2, . . ., Ag_M\}$ which each is willing to be an expert on one of the main classes. We try to make any Agent, $Ag_i$ to be an expert in deciding over one and only one of the main classes of the system, annotating the known features of their main classes, according to the probability of those features representing the concepts for incoming queries. These agents have collaboration under the name of MAS and have the relation called "Peer" with eachother. It's somehow clear that at most attention, yet these main classes may have something in common. But this fact is not an obstacle to our approach while the expertism of any agent over its own class is important here, and is a triumph over such hindrances.

To moderate the action of dispatching the incoming query among agents/classes, we utilize a more general and abstract agent, called the CenterAgent and has the relation called "Pair" with any agent representing a main class. This entity of our system maintains an abstract information of all other agents of the system at any time, and due to this information, decides to forward the query to a subset of all its agents. The task of dispatching is held under some value of probability, called the "degree of confidence" which shows to what extents our CenterAgent thinks the object is for the pair agent to decide upon. It then incepts the results of all its pair agents, does a mixing job on these results with regard to the degree of confidence and gets a perception on in which class or classes would the incoming object a consonant be. This result is outputted as a set of concepts with the most probable posse. Figure 1 shows how this system works when you look at it from outside.

Figure 1. is the big picture of the system being discussed

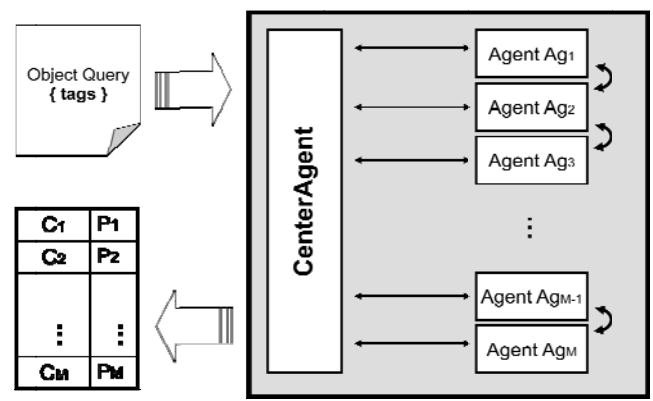

Another important matter in our approach is that each agent, having a set of features for its corresponding main class, is able to revise the probabilities of features expressing the concept of that main class in time with the help of incoming queries. These queries are also utilized as a shape of suitable dataset to learn the agent on features related to its main class.

A talk on more details comes next, and the discussion of systems operability and performance after.

### III. More Discussion

#### A. Before Running the System

Here we provide the procedure of our system. At first, we assume that our system is equipped with some base classes. These main classes are gathered and approved by a group of experts before the system initialization, and then are fed into the kernel of system. Each main class represents a somehow abstract concept which is named here as a SuperConcept. This SuperConcept is the parent of a group of more detailed concepts in the discussed criteria, which these concepts are the children of the main classes. This is as we are talking about an abstract agent like CenterAgent holding the control of some more concrete agents like its pair decision-expert agents. The hierarchy is shown in Figure 2.

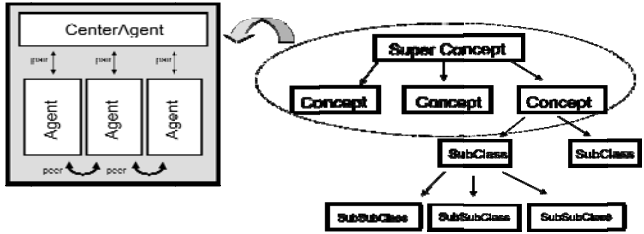

Figure 2. shows the hierarchy of concepts and the relation of this hierarchy to the system model being discussed

Each of these abstract main classes holding the Concept of their own, are modeled with a group of features, while a concept is a group of objects having the common characteristics of their own in general. These characteristics are here named as features and are held in a feature set. The hierarchical representation we are using here lets us provide any node in the lower levels of the hierarchy as a SubConcept of the SuperConcept, holding a subset of the base features representing the parent SuperConcept.

These base features we are discussing here are also fed into our system before its initialization with help of the experts running the system. A set of all base features for all M classes are reserved into the CenterAgent and some M disjoint subsets of our CenterAgent's base features into any main class agent, pointing the concepts of that class.

Before discussing about the starting of the system, it is essential to explain more on some idea we are utilizing. We define two thresholds, after Williams [1] for governing the probabilities of each class's features, being held in each agent. Any feature including the base features of subjective ideas from experts at the beginning of system running, is placed into a region above the first threshold. We name this region as the K region as [1]. Any feature between the two thresholds is placed in yet another region, named as the M region. Any feature below the second threshold is placed into a D-region. The K-region features have the most probability to show the concepts under a main class, while being revised to give the best performance of classification to the system. The M-region features have a lower probability showing the main class, but are authorized to either enter the K-region or even leave the feature set for the D-region. The D-region features are those who may not reflect any property of the class they are in, but are reserved in the agent feature collection in order to be kept track –maybe the dispatching was false over a long period of time- and to be allowed to yet come back to the regions of impacting on generating the result. Figure 3 shows one pair agent's feature set.

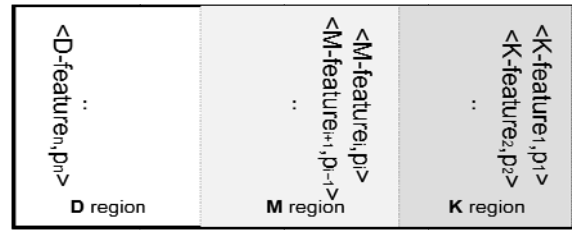

Figure 3. Figure (3) is the representation of an agent's feature collection, including the K, M and D-region features with their probabilities

This all means that each agent of a main class has a set of K-region features with upper threshold K-region probability and a set of M-region features with probabilities between the K and M thresholds. These feature sets are the main help on generating and sending results back to the CenterAgent on its dispatched query. Yet another set of all these K-region features of all classes are gathered together in the CenterAgent to help it decide about the dispatching task of any later incoming query. This is shown in Figure 4 below.

Figure 4. show the relation between CenterAgent and its pair agents on the feature collection

### B. Running the system

When a newly come object arises to our system, a pre-process is done to extract the collection of features in the object, which are named as tag collection. This queried object is then forwarded to CenterAgent as the collection of extracted tags. The CenterAgent, now reads the queried object using its collection of tags and based on its abstract knowledge of its pair agents –which is the base feature set discussed above to be held in the kernel of this system- and their expertism, decides to dispatch this query in a standard view including its extracted tags to one or some pair agents. As once mentioned, this dispatching task is under a degree of confidence which is the degree our CenterAgent thinks the dispatching is done truly for any pair agent. These pair agents would be as a consultant to our CenterAgent on deciding about the classification task. Each pair agent $Ag_i$ now viewing over its set of features and probabilities of expertism, say $\{<f_j, p_j>|\ f_j \in K_i \cup M_i\}$, searches for mapping features to some concept with a probability and packages this as a result of the incoming query. This package is sent back towards the CenterAgent. Our CenterAgent sorts these packages of results from its pair agents and generates a vector of outgoing output looking at its degree of confidence for any result coming from any pair agent. This vector consists of main classes and the likelihood of the incoming queried object representing any of them. This vector should be what is seen as the result of the system but its user.

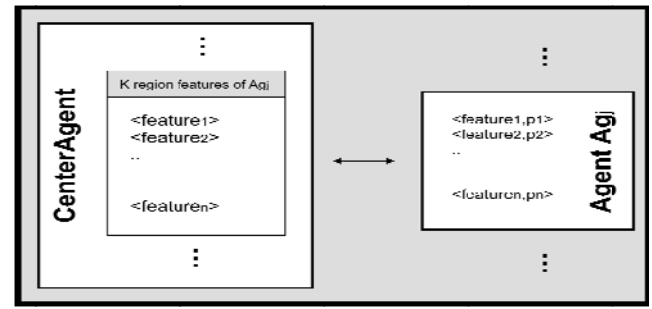

## C. Learning more on each agent's feature probabilities

Our object-classification multi-agent system does its task with the help of the collection of features and probabilities, in which it stores some K, M and D region features, representing the subconcepts of its own main class. While having these features on the hierarchy of probabilities, any willing-to-be-expert agent meets some dispatched queries from the CenterAgent. These objects, for sure include some several features, here tags, which may be a member of any main class collection or not. This may lead to have a feature dataset for any agent to individually learn and revise its feature collection and their corresponding probabilities due to the incoming dataset. Whenever a new object consisting of several tags is passed to an agent, it is going to keep track of the tags inside the query and their density over time. For this purpose, any expert agent –in addition to its collection of features and corresponding probabilities- is equipped with a time-based memory, named the "Time-Interval Memory" to keep track of most recently incoming features. This time-interval memory is shown in Figure 5. Here the agent can discover any needed feature for describing its own concepts over the whole features of the queries being passed to it. If a feature $f_j$ is coming over and over for $Ag_i$ and it is in the base features of it, then the probability that this feature is more helpful for describing $C_i$ concepts is relevant. Unlike it is cool to lower the impact of a feature on decision making tasks which we are not so sure about it repeating in $C_i$ concepts.

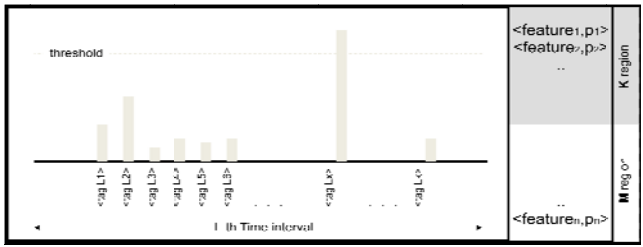

Figure 5. Figure (5) shows the structure of any pair agent on holding the feature collection besides its time-interval memory

## D. Learning best features by any agent

The above told tag collection is also a good dataset for agents to collaborate within and do some extra learning over features with the help of the peer agents. Along with the process of object classification, any agent has a correlation with its peer agents, trying to improve the expertism of the MAS by consulting on whether they know anything about its now deciding features.

For this purpose, again we utilize the time-interval memory keeping track of most recently incoming features. Almost these tags are parts of a query which is dispatched from CenterAgent, there may come some features which could be helpful to Agi in its future decision making processes, i.e. tags which could be a base feature of Ci but is not mentioned at the beginning by the expert team. Here Agi needs to consult with its peer agents. Simply talking about this procedure, Agi consults on the non-familiar features which are repeated over its memory time slice over and over.

If this process of consulting leads to this perception that this newly-come feature should be added to Agi's feature set, then there has to be some raise/reduce modification on peer agents' feature probabilities.

Let us suppose that Agi has found some new feature, say fj, in its time-interval memory interesting for itself to add to its feature set. Now Agi adds this feature to its feature set with an appropriate probability, let to be the floor probability of its M-region. As it finds the feature more interesting, it adds the probability of this feature to some extent, let us suppose α. This is done over and over until this feature reaches to the border of the M and K regions, which is the first defined threshold. This procedure is called the raise procedure. Now that Agi has found out that this feature has a lot in common with its K-region features and could be a base feature for the main class it is representing, it would like to add the feature probability again and put it up to its K-region. But this may cause a problem while we have governed the K-region features of all classes to be disjoint subsets of all-features set (being reserved in the CenterAgent), and it is possible for fj to be a member of some other agents' K-region feature sets. In order to solve this obstacle, we provide an approach as below:

"Whenever an agent Agi reaches to the point where it finds a feature fj suitable to enter its K-region, let the agent consult with its peer agents on whether they have such a feature in their K-region. If not so, add fj to the feature sets of Ci with some appropriate probability, let to be the floor probability of K-region. But in the case of having the contradiction within the peer agents, let Agi reduce the probability of fj for the peer agents having it in their K-region to some extent, let us again suppose α. This procedure is called the fall procedure. Let Agi know this fact that no peer agent has fj in its K-region when the fall procedure comprises the contradiction. Let Agi add fj as above, now."

The more interesting this feature is for Agi, the more it calls for all peer agents for their advisory sight, and the most it causes the α-value probability reduce happen. This is done until we can have yet again disjoint sets of base features for each class by adding fj to Agi's class feature set.

## E. Differences with prior attempts on the topic

Several attempts have been made on concept learning and object classification, of which the one that is so important in our point of view is Williams' [1]. This importance shows itself in the list of its citations! So we provide some outstanding differences in a comparison between our approach and the ones based on [1] and Williams himself.

Williams' work is based on features, as ours. But he uses the K, M and D regions for deciding about the addition of concepts. This is the first point of disagreement between the two approaches. We utilize these regions to moderate the features in an agent's feature collection. This moderation is for deciding about which features to let in deciding about any incoming query or easier to say, to decide about the addition of features. The next is Williams uses features to generate several rules and then uses these rules to decide. But we believe that the performance of such systems is directly dependant to the number of rules we use. Better to say, each time we ask something in our system, we are using some resources. So it is better to lower the number of these queries. Another matter is that the probabilities which Williams use are put over the rules that he uses. In our approach, the rule set is somehow constant and the decision making task is due to the feature collection and their corresponding probabilities. It is once told above that this feature-probability collection is allowed to shrink or expand during system working time.

As a big picture to get the differences, Williams uses some agent to collaborate about learning some concepts with questioning eachother whether anyone knows anything about the mentioned concept. We are using a MAS with several expert agents, each in some realm of knowledge, to decide on classifying one object into one or some of the realms, known as concepts. This is done by utilizing the implicit knowledge of our expert agents. This knowledge is represented as a collection of coupled feature-probability metadata.

*F. Fault Tolerance, Performance and Operability*

It is clear from the above section "Learning more on each agent's feature probabilities" that if some feature $f_j$ is wrongly supposed to be added to some class's feature sets and it is inserted to the M-region for that reason-and even increased during some interval of false tag mentioning-, after a period of time that it is not mentioned in its forwarded objects by the CenterAgent, we would have the system automatically lower its probability to lower M-region and may also to D-region not to have bad impact on our system. This fact exists in the steady state for the wrong features that some expert would falsely put into some agent's body.

Another matter is the performance of this system, trying to make its agents expert in some fields. As time goes by, the features of a same kind and similar –not in essence but in concept recognition- gather together in collections, showing the concepts they are the representative to identify with. Besides, the reserved collection of all the disjoint class features sets in the CenterAgent makes the system lower the message passing rate among the pair and peer agents after a period of system execution. This means that the CenterAgent has to consult with fewer consultants and even get a better result and it is indeed an economical thrift in the realm of decision making. Of course that it will not appear when the class numbers are in the order of 10 or 100. But suppose that we are willing to decide upon 100000 or 1000000 classes to correlate some objects into them. Using the methods based on [1] and of course itself, any agent should consult to any other agent on knowing anything about the decision matter. Unavoidably a great extent of this message passing is for nothing while there are so many agents that are not aware of what the concept is or what objects point to it. Using our approach in a MAS, the CenterAgent consults only to some agents -of an order of for example 50 or 100- whether they could help the decision making problem. It is discussed before that these agent are the most probable agents on most probable concepts that the pointed object could be a match for. We believe that it is clear that such a system is better in performance in comparison to its parents. We also believe that the operability of such approach in any realm of application is fully applicable and also implementable.

CONCLUSION AND FUTURE WORK

Here we addressed a method of concept learning in a Multi Agent system, in which several agents are labeled to be an expert in a specific domain of itself. This is supposed to be a good way for object classification -into some known main classes- in almost any application domain. Along with this going on, we would like to let our system add needed (willing to be expert) agents for any newly accepted main concept classes. Another work is to revise the likelihood of concept features, based on their agent-rules along the time. We would like to mention these tasks in near future.

REFERENCES

[1] A. B. Williams: Learning to Share Meaning in a Multi-Agent System; Autonomous Agents and Multi-Agent Systems, 8, 165-193; 2004 Kluwer Academic Publishers, Manufactured in Netherlands.

[2] L. Gravano, P. G. Ipeirotis, M. Sahami: QProber: A System for Automatic Classification of Hidden-Web Databases

[3] M. Afsharchi, J. Denzinger, B.H. Far: Enhancing Communication with Groups of Agents using Learned Non-Unanimous Ontology Concepts; Web Intelligence and Agent Systems 3 (2007) 1-3; IOS Press;

[4] B. Coppin: Artificial Intelligence Illuminated; Jones and Bartlett Publishers 2004; Sudbury, Massachusetts; pages 543-546.

[5] M. Afsharchi: Ontology-Guided Collaborative Concept Learning in Multiagent Systems; PhD Thesis; DEPARTMENT OF ELECTRICAL AND COMPUTER ENGINEERING, UNIVERSITY OF CALGARY; pages 13-15.

[6] J. M. Vidal: Fundamentals of Multiagent Systems; 24$^{th}$ August 2007; page 9.

[7] N. J. Nilsson: Artificial Intelligence, a New Synthesis; Morgan Kaufmann Publishers Inc., 1998.

[8] S. J. Russell, P. Norvig: Artificial Intelligence, a Modern Approach, 2$^{nd}$ Edition; Prentice Hall Series in AI, 2002.

[9] M. Afsharchi, B.H. Far, J. Denzinger: Ontology-Guided Learning to Improve Communication between Groups of Agents; Proceedings of the 5$^{th}$ International Joint Conference on Autonomous and Multi-Agent Systems, 923-930; Hakodate, Japan 2006.

[10] D. Koller, M. Sahami: Hierarchically Classifying Documents using few words.